\documentclass{article}

\usepackage{PRIMEarxiv}
\usepackage{algorithm}
\usepackage{algpseudocode}
 \usepackage{multirow}
\usepackage{amsmath}
\usepackage[utf8]{inputenc} 
\usepackage[T1]{fontenc}    
\usepackage{hyperref}       
\usepackage{url}            
\usepackage{booktabs}       
\usepackage{amsfonts}       
\usepackage{nicefrac}       
\usepackage{microtype}      
\usepackage{lipsum}
\usepackage{fancyhdr}       
\usepackage{graphicx}       
\graphicspath{{media/}}     

\title{Application Research on Real-time Perception of Device Performance Status
}

\author{
  Zhe Wang \\
  ByteDance Inc. \\
  Beijing, China \\
  \texttt{wangzhe.gigi@bytedance.com} \\
   \And
  Zhen Wang \\
  ByteDance Inc. \\
  Beijing, China \\
  \texttt{wangzhen3560@bytedance.com} \\
  \AND
  Jianwen Wu\\
  ByteDance Inc. \\
  Shenzhen, China \\
  \And
  Wangzhong Xiao\\
  ByteDance Inc. \\
  Shenzhen, China \\
  \And
  Yidong Chen \\
  ByteDance Inc. \\
  Shenzhen, China \\
  \And
  Zihua Feng \\
  ByteDance Inc. \\
  Shenzhen, China \\
  \AND
  Dian Yang \\
  ByteDance Inc. \\
  Singapore City, Singapore \\
  \And
  Hongchen Liu \\
  ByteDance Inc. \\
  Beijing, China \\
  \And
  Bo Liang \\
  ByteDance Inc. \\
  Beijing, China \\
  \And
  Jiaojiao Fu \\
  ByteDance Inc. \\
  Beijing, China \\
}

\begin{document}
\maketitle

\begin{abstract}
In order to accurately identify the performance status of mobile devices and finely adjust the user experience, a real-time performance perception evaluation method based on TOPSIS (Technique for Order Preference by Similarity to Ideal Solution) combined with entropy weighting method and time series model construction was studied. After collecting the performance characteristics of various mobile devices, the device performance profile was fitted by using PCA (principal component analysis) dimensionality reduction and feature engineering methods such as descriptive time series analysis. The ability of performance features and profiles to describe the real-time performance status of devices was understood and studied by applying the TOPSIS method and multi-level weighting processing. A time series model was constructed for the feature set under objective weighting, and multiple sensitivity (real-time, short-term, long-term) performance status perception results were provided to obtain real-time performance evaluation data and long-term stable performance prediction data. Finally, by configuring dynamic AB experiments and overlaying fine-grained power reduction strategies, the usability of the method was verified, and the accuracy of device performance status identification and prediction was compared with the performance of the profile features including dimensionality reduction time series modeling, TOPSIS method and entropy weighting method, subjective weighting, HMA method. The results show that accurate real-time performance perception results can greatly enhance business value, and this research has application effectiveness and certain forward-looking significance.
\end{abstract}


\section{Introduction}
The efficient and stable operation of mobile devices plays a crucial role in the continuity and effectiveness of production and services in today's highly industrialized and automated era. The dynamic performance state of the device directly affects its work efficiency, reliability, and product quality. Therefore, accurately sensing the real-time performance state of the device and predicting its performance change trend has become a key factor in ensuring the normal operation of the device, optimizing experience strategies, and enhancing business value and production efficiency.

Due to the high precision, intelligence, and other characteristics of mobile devices, limitations such as computing resources, battery life, environmental temperature, and network conditions continue to pose a major challenge to achieving high-quality, smooth, and stable operation. Existing device performance evaluations only treat devices with uniform hardware configurations under the same model equally. However, in differentiated usage scenarios, the degree of device performance degradation varies, which limits the operational space for reducing maintenance costs for device performance stability, fine-tuning performance experience strategies, and predicting performance change trends. For example, a phone that has been purchased for one year and has been used continuously for eight hours will show significantly different performance data in operation compared to a brand-new phone of the same brand, model, and production period that has just been purchased and started to be used.

In order to improve the user experience of mobile devices, a new type of performance real-time perception model has been researched and developed. It is based on real-time performance indicators on mobile devices, using feature engineering methods such as dimensionality reduction and correlation analysis to construct an objective weighted multi-level evaluation model and time series model to evaluate and predict device performance status. The evaluation results can be used as a basis for production and business means adjustment, to reasonably intervene or optimize device performance, and achieve the best balance between device usage experience and real-time system performance status.

The research on real-time perception model of performance status mainly includes three parts: extraction and construction of performance characteristics, multi-level evaluation model with weighting, and time series-based prediction model.

During the feature construction process, directly applying features that affect the performance state of mobile devices can inevitably lead to overfitting during model training due to the diversity and complex attribution of the distributed features. Using brute force methods to directly reduce multiple features can lead to defects such as reduced anti-interference ability, decreased sensitivity, and redundant important information. Therefore, in addition to conducting feature correlation analysis and selecting appropriate features using principal component analysis, it is also necessary to conduct descriptive time-series analysis modeling of performance feature data with time-series characteristics, and aggregate and map them into device profile-like features to improve model input. In the multi-level evaluation model construction process, the comprehensive performance of the feature values of each performance module and the influencing factors within each module can more comprehensively describe the real-time performance status. Therefore, a hierarchical evaluation model is used to progressively evaluate each performance module and obtain a comprehensive performance status evaluation result. In addition, objectively determining the degree of influence of each feature factor on device performance and assigning weights to input evaluation models is the difficulty of real-time performance perception algorithms. The optimal and worst solutions are determined using the superiority and inferiority distance method\cite{papathanasiou2018topsis}, and sorted based on distance. The entropy of the features is calculated\cite{1965Lectures} to objectively assign weights based on the relative degree of change in the feature's own value, which solves the difficulty of multi-level performance state evaluation modeling. In the final time series prediction model, considering that the performance state perception result at each instant is affected by the real-time performance feature value changes, it is aggregated into a non-stationary sequence of discrete numerical values. Accurate prediction of the next performance state of the device requires smoothing of the noise data and short-term fluctuations contained therein, and increasing sensitivity while reducing lag. Compared with ordinary weighted moving average, the Hull moving average (HMA)\cite{2018OPTIMISING} is fast and smooth, eliminates lag while providing a smooth curve to reflect the trend of performance state changes.

\section{Architecture}
\label{sec:Arch}

\begin{figure}
  \centering
  \includegraphics[width=1\textwidth]{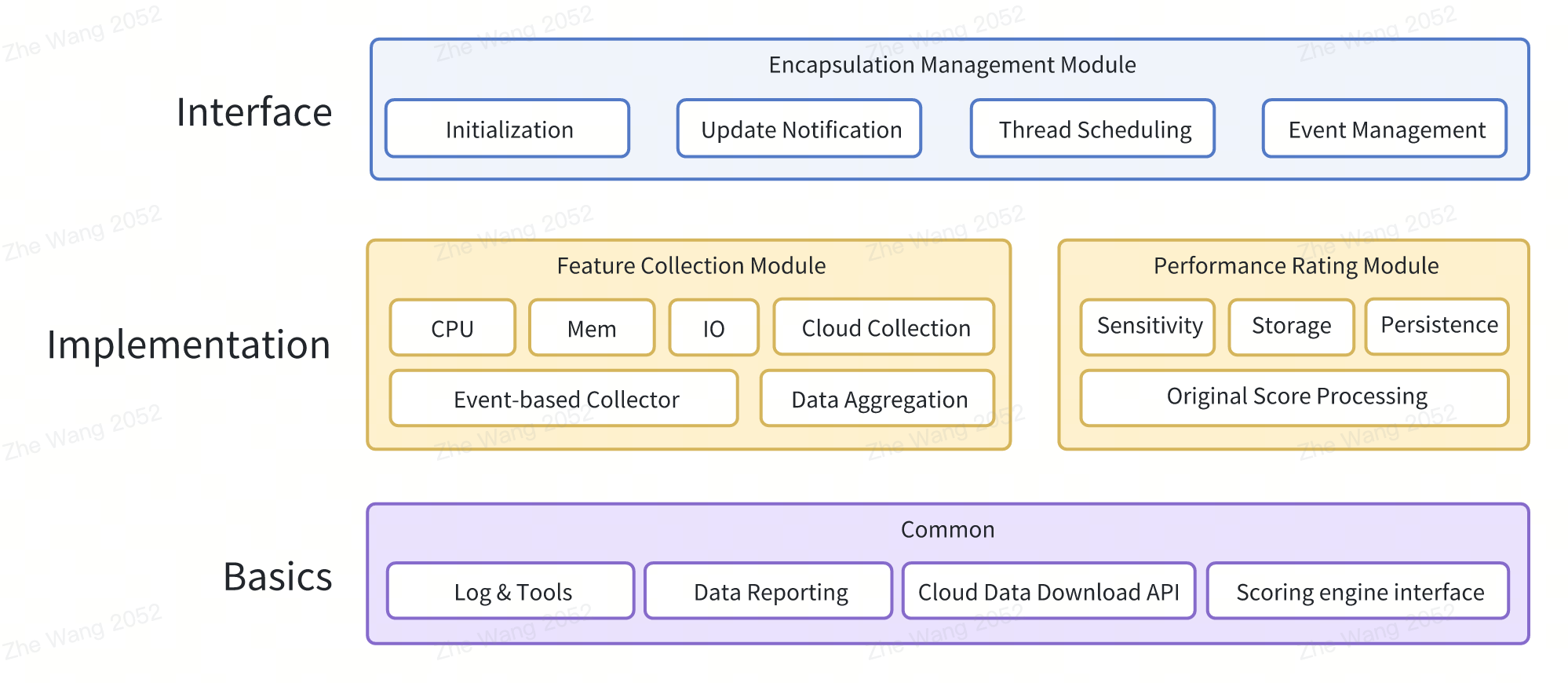}
  \caption{Real-time Performance Evaluation Service Framework for Devices.}
  \label{fig:1}
\end{figure}

The overall framework for real-time evaluation of device performance status is implemented on the device client, and is accessed through a common library using the media used in this study, namely the video playback software Douyin. This framework and model can be reused in any application software. The main functional modules are shown in Figure \ref{fig:1}.

The main functional modules include three parts:
\begin{enumerate}
    \item Encapsulation management module: provides a simple and easy-to-use interface for upper layers, realizes initialization and configurable capabilities, manages internal modules, implements thread scheduling and runtime management.

    \item Feature Collection Module: Implements the ability to collect local and cloud features (requiring time series aggregation and portrait label extraction) and is responsible for aggregating feature items and providing a unified read interface.

    \item Performance rating module: implements the logic module for real-time calculation and management of ratings, calls the algorithm engine to obtain raw real-time evaluation scores, and implements state evaluation with multi-level time-domain sensitivity, as well as storage logic and persistent management.
\end{enumerate}

\begin{figure}
  \centering
  \includegraphics[width=1\textwidth]{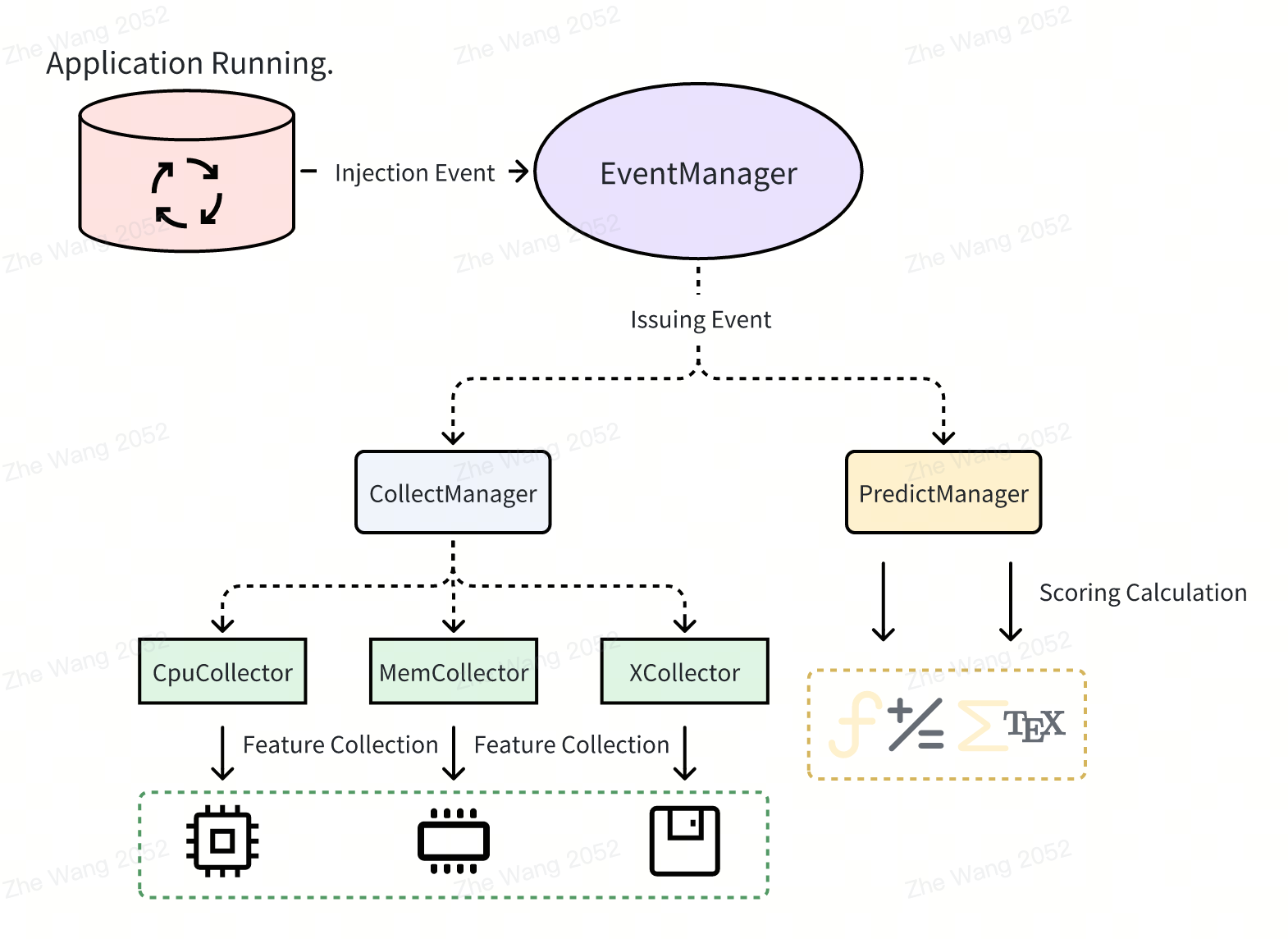}
  \caption{Real-time evaluation service for device performance status driven by events.}
  \label{fig:2}
\end{figure}

The event-driven service is an important mechanism within the framework, which realizes the trigger timing and feature collection timing of real-time evaluation scoring. The timing can be flexibly configured, including built-in common trigger event points, and the ability to expand capabilities under personalized requirements can also be achieved by adding custom events externally, see that in Figure \ref{fig:2}.

The injection of event-driven is applied at key nodes by injecting event-driven into the framework. For example, startup and playback events have globally unique string names and can carry parameters. The startup event can carry startup duration, and the playback event can carry playback-related information. When integrating with Douyin, some common events are already built-in, and the framework also provides interfaces for business partners to add custom events. The distribution of event-driven is achieved through registration, and events can be distributed to the main internal modules' drivers. The feature collection module and performance scoring module both listen for events that trigger their own operation. The event that triggers the scoring operation is configurable, and the event that triggers the feature collection is determined by the collection strategy, which external parties do not need to be concerned about. For example, the startup feature is collected at startup, and the CPU and memory features are collected before scoring.

\section{Algorithm}
\label{sec:Algo}
The algorithm mainly consists of two parts: a dynamic multi-level fuzzy evaluation model of the device performance based on TOPSIS and entropy weight method, and the performance score sequence smoothed within a time window.

\subsection{Real-time Performance State Evaluation Based on TOPSIS + Entropy Weight Method}

The Technique for Order Preference by Similarity to Ideal Solution (TOPSIS) was first proposed by C.L. Hwang and K. Yoon in 1981. It ranks a set of objects based on their proximity to the ideal values of a finite number of evaluation criteria, providing an overall assessment of the object under a certain set of criteria. TOPSIS is suitable for multi-objective decision analysis, as it eliminates dimensional influence through normalization and positive transformation of evaluation criteria, making full use of the original data set with large distribution differences and no distribution restrictions. The results accurately reflect the differences between evaluation criteria and the comprehensive impact on the evaluation object.

The algorithm process of TOPSIS is as follows:
\begin{enumerate}
    \item Positive Transformation of original data matrix:

    Convert all evaluation indicators (i.e. features) to maximization type indicators, which gradually approach the optimal target value as the indicator value increases; forward normalize the original dataset matrix. The conversion functions have the following forms:
    \begin{enumerate}
        \item  Converting intermediate indicators to maximal indicators:
  
  $M = max\{|x_i-x_{best}|\}, x_i = 1-{\frac {|x_i-x_{best}|} M}$
        \item  Converting a minimal indicators into a maximal indicators: $max-x$
        \item  Converting interval indicators to maximal indicators:
        
  $M = max\{a-min\{x_i\}, max\{x_i\}-b\}, x_i = \begin{cases}
    1-\frac {a-x}M,&x<a \\
    1  ,&a\leq x\leq b \\
    1-\frac {x-b}M,& x\geq d
  \end{cases}, \{x_i\}, [a,b]$
 \end{enumerate}
 \item Standardization of Positive Matrices:
 The dimensions and distributions of each indicator differ, so standardization calculations are performed on the positively transformed data matrix to eliminate the impact of indicator dimensions on evaluation. Assuming there are $n$ target objects to be evaluated, and $m$ evaluation indicators make up the positively transformed matrix: \begin{equation}
    X=\begin{bmatrix}
   x_{11} & x_{12} ... x_{1m} \\
   x_{21} & x_{22} ... x_{2m} \\
.
.
.\\
x_{n1} & x_{n2} ... x_{nm}
\end{bmatrix}\end{equation} 
The matrix obtained by standardizing each of its elements is denoted as$Z$. The standardization method for elements is: \begin{equation}z_{ij} = {\frac {x_{ij}} {\sqrt{\displaystyle\sum_{i=1}^n
x_{ij}^2}} }\end{equation}
\item  Calculate the advantages and disadvantages score.

By calculating the distance between the evaluated object and the best and worst targets, the score of the evaluated object is obtained. The standardized matrix is as: 
\begin{equation}
Z=\begin{bmatrix}
   z_{11} & z_{12} ... z_{1m} \\
   z_{21} & z_{22} ... z_{2m} \\
.
.
.\\
z_{n1} & z_{n2} ... z_{nm}
\end{bmatrix}.
\end{equation}
Record all the optimal indicators combination as the optimal goal $Z^+$: \begin{equation}Z^+=(Z_1^+,Z_2^+,...,Z_m^+)\end{equation}
\begin{equation}Z^+=(max\{z_{11},z_{21},...,z_{n1}\}, max\{z_{12},z_{22},...,z_{n2}\}, ..., max\{z_{1m},z_{2m},...,z_{nm}\})\end{equation} 
Record all the worst indicators as the optimal and worst targets $Z^-$: \begin{equation}Z^-=(Z_1^-,Z_2^-,...,Z_m^-)\end{equation}
\begin{equation}Z^-=(min\{z_{11},z_{21},...,z_{n1}\}, min\{z_{12},z_{22},...,z_{n2}\}, ..., min\{z_{1m},z_{2m},...,z_{nm}\})\end{equation}. 

The distance between the $i$th evaluation object and the optimal target is:
\begin{equation}D_i^+=\sqrt{\displaystyle\sum_{j=1}^m (Z_j^+-z_{ij})^2}\end{equation}  
The evaluation object and the optimal target is:
\begin{equation}D_i^-=\sqrt{\displaystyle\sum_{j=1}^m (Z_j^--z_{ij})^2}\end{equation} 
The $i$th evaluation object's non-normalized score for overall performance is obtained 
\begin{equation}S_i = {\frac{D_i^-}{D_i^++D_i^-}}\end{equation} 
When $0\leq S_i\leq 1$, the larger the value of $S_i$, the larger the value of $D_i^+$, which means it is closer to the optimal target value.
\end{enumerate}

The above is the evaluation score of the comprehensive evaluation object under the condition that the weight of each index is the same. In reality, it is often necessary to analyze the weight and PCA principal component analysis\cite{2014A} of indicators with different degrees of discreteness or indicators that are difficult to obtain and have correlation and collinearity problems. In information theory, entropy is a description of the discreteness of indicators and can measure their uncertainty. Generally speaking, the larger the amount of information, the smaller the information entropy value, and the greater the discreteness of the indicator and the smaller the uncertainty, which can be understood as the greater the impact (i.e. weight) of the indicator on the comprehensive evaluation object. Therefore, the entropy weight method can guide the determination of the objective weight of indicators by calculating the relative degree of change of each indicator's overall impact on the evaluation object.

The algorithm process of entropy weight method is as follows:
\begin{enumerate}
 \item Standardization of Positive Indicators: Each index in the corresponding positive normalization matrix $X$ is standardized using \begin{equation}x_{ij}'=[\frac {x_{ij}-min(x_{1j},x_{2j},...,x_{nj})}{max(x_{1j},x_{2j},...,x_{nj})-min(x_{1j},x_{2j},...,x_{nj})}]*100\end{equation}
 \item Calculate the impact of the $j$th indicator on the $i$th evaluation object, that is, the weight of the indicator: \begin{equation}p_{ij}=\frac{X_{ij}}{\displaystyle\sum_{i=1}^nX_{ij}} ,(i=1,2,...,n, j=1,2,...,m)\end{equation}
 \item The entropy value for the $j$th index is calculated as: \begin{equation}e_j=-k\displaystyle\sum_{i=1}^np_{ij}\ln(p_{ij})\end{equation} 
 which $k>0$ and $k=\frac1{\ln(n)}$, $e_j\geq0$;
 \item The coefficient of variation for the $j$th indicator is calculated as: \begin{equation}g_j=1-e_j, (j=1,2,...,m)\end{equation}
 \item Calculate the weight coefficient of the $j$th indicator: \begin{equation}w_j=\frac{g_j}{\displaystyle\sum_{j=1}^mg_j} ,(1\leq j\leq m)\end{equation}
 \item The score for the non-normalized evaluation of the $i$th evaluation object is calculated as follows: \begin{equation}s_i={\displaystyle\sum_{j=1}^mw_j}*p_{ij}, (i=1,2,...,n)\end{equation}
\end{enumerate}

By integrating the above algorithm process, a comprehensive evaluation score of the evaluated object in the overall set can be obtained. Based on TOPSIS combined with entropy weight method, it can deeply understand and reflect the degree of dispersion and discrimination ability of each evaluation index, and then determine the global influence coefficient of the index on the evaluated object. It is an objective and comprehensive weighting method with high reliability and accuracy. However, this method also has its flaws. Due to the single algorithm process, it does not consider the correlation and hierarchical relationship between indicators. In the application scenario, business experience guidance is needed to assist in determining the value of indicator weights, otherwise the weights may be distorted due to differences in application scenarios. In addition, both methods only calculate the evaluation indicators, and have a strong dependence on the sample data of the indicators. If the sample data is dynamically changing, the indicator weights will fluctuate to a certain extent.

In response to the shortcomings of the algorithm plan mentioned above, we have incorporated an analysis of the correlation between indicator features in our research. Additionally, we have conducted multi-level evaluation calculations\cite{1994A} and principal component analysis on indicators with hierarchical relationships. Furthermore, we have designed device profile features that describe the time sequence to ensure the stability of indicator sample data within a certain time window and to determine the stable and smooth indicator weights under the time dimension.

\subsection{HMA Performance State Comprehensive Perception}

In the research of comprehensive prediction with dynamic performance awareness, in order to provide diversified sensitivity (real-time, short-term, long-term) perception ability and assist in eliminating abnormal mutations in indicator data and evaluation scores, it is necessary to smooth the performance evaluation results. That is, in addition to real-time perception of device performance status, the average performance status of devices in different time windows should also be comprehensively evaluated.

Due to the fact that the performance indicator data of the equipment does not change smoothly within the specified time window, extreme values, null values, and other situations that require preprocessing exist, and there are also error values that are difficult to monitor and objectively distinguish. In addition, the performance of the equipment also depends on a series of objective factors such as environmental changes, usage patterns, and frequency. Therefore, in the process of performance indicator collection and performance status evaluation score calculation, there is a chance of generating noisy data or weight values and score values with large fluctuations. To compensate for such defects, in the short-term and long-term comprehensive performance status evaluation process, the study has adopted Hull Moving Average to quickly smooth the instant performance score curve within the time window and accurately fit the trend of performance status changes.

The Hull Moving Average (HMA) is a type of moving average proposed by Alan Hull in 2005. It uses weighted calculations to emphasize recent data changes. HMA improves fast smoothing capabilities while eliminating the lag of other moving averages, minimizing short-term noise and improving accuracy. The pseudocode for HMA calculation method is:

\begin{algorithm}
\caption{Hull Moving Average Algorithm}\label{alg:hull}
\begin{algorithmic}[1]
\Require $score\_series: List[float]$, $lookback: int$
\Ensure $hma\_series$
\State $wma\_full=WMA(lookback)$ of $score\_series$
\State $wma\_half=WMA(lookback/2)$ of $score\_series$
\State $raw\_hma=(2*wma\_half)-wma\_full$
\State $wma\_sqrt=wma(sqrt(lookback))$
\State Use $wma\_sqrt$ smooth the origin $raw\_hma$
\For {$i$ in $raw\_hma$}{

{$hma=wma\_sqrt*i$}}
\EndFor
\State $hma/((lookback**2+lookback)/2)$
\end{algorithmic}
\end{algorithm}

\begin{figure}
  \centering
  \includegraphics[width=0.8\textwidth]{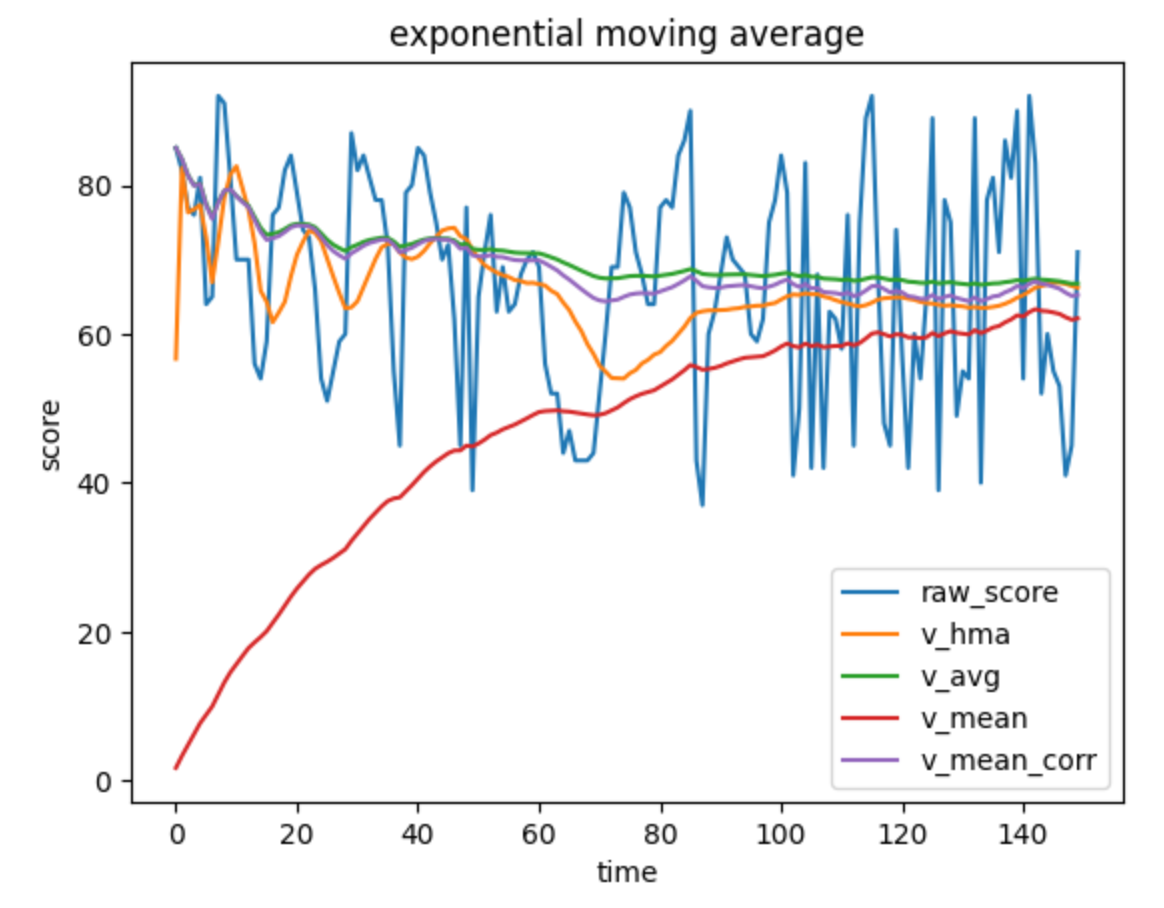}
  \caption{Smoothing results of different moving averages in the original performance dynamic scoring sequence.}
  \label{fig:3}
\end{figure}

By comparing the experimental sliding average line chart in Figure\ref{fig:3}, it can be seen that in a 150-point real-time device performance status scoring sequence (raw\_acore), the simple sliding average (v\_avg) remains stable during the process of increasing and decreasing the window, with the most severe lag, that is, in the trade-off between smoothness and sensitivity, it chooses smoothness and loses sensitivity, and does not have good capturing ability for the trend of performance scores; the weighted sliding average (v\_mean) has a greater dependence on the selection of weight factors and window length, which requires subjective judgment and selection, and also has a certain lag, with poor sensitivity to changes in perception scores; the corrected weighted sliding average (v\_mean\_corr) has slightly optimized the lag and sensitivity of the simple sliding average, and has reduced the larger deviation in the initial stage compared to the ordinary weighted sliding average, but this method still cannot meet the expected sensitivity to perceive changes in scores; the Hull sliding average (v\_hma) performs the best compared to other methods, with the biggest feature being a significant reduction in lag, while improving sensitivity and effectively improving the smoothness of the moving average.

By using HMA calculation, the average evaluation results of device performance status within a time window, periodic changes, and the changing trend of performance status at a certain moment can be obtained. In the research of prediction, because the performance status of the next moment depends on the evaluation score of the current moment and the past several moments of the device's performance status, a self-regression model is selected.

\subsection{Performance state prediction based on ARIMA time series}

The ARIMA (Autoregressive Integrated Moving Average model)\cite{2000A} provides the ability to predict and analyze non-stationary time series, modeling non-seasonal time series based on historical values and historical forecast errors.

The modeling process of ARIMA can be summarized as the calculation process of its three hyperparameters $p$, $d$, and $q$. $p$ is the order of the AR autoregressive model, which predicts the current value based on the previous $i$ historical values. The definition of the AR model is $X_t = c + \sum_{i=1}^p \varphi_i X_{t-i}+ \varepsilon_t$, which means that the predicted current value is the sum of a linear combination of one or more historical values, a constant term $c$, and a random error $\varepsilon_t$. $d$ is the minimum order of differencing required to make the time series stationary. Non-stationary series can be transformed into stationary ones by differencing, but a larger order of differencing will cause the time series to lose autocorrelation, making the AR autoregressive model unusable. $q$ is the order of the MA moving average model, which predicts the error between the previous $i$ historical values and the current value. The model definition of MA is $x_t = \mu + \varepsilon_t - \theta_1 \varepsilon_{t-1} - \theta_2 \varepsilon_{t-2} - ... - \theta_i \varepsilon_{t-i}$, where $\mu$ is the mean of the sequence, $\theta_1$,..., $\theta_i$ are parameters, and $\varepsilon_t$,..., $\varepsilon_{t-i}$ are all white noise.

After performing differential processing on the score sequence obtained from 50-300 consecutive evaluations of a certain device's performance, a stable time series is obtained by determining the differential order $d$. By using the above equation, the order $p$ of the AR model and the order $q$ of the MA model can be determined, and an ARIMA model can be constructed. The $p$ and $q$ parameters of the model can be verified by using information criteria such as AIC, AICc, and BIC to determine the order scheme, with smaller information criteria indicating better parameter selection. 

\section{Experimental Analysis}
\label{sec:Exp}

The experimental process of the real-time perception model for performance status mainly includes two aspects: extraction and construction of performance characteristic indicators, and verification of multi-level evaluation model with weighting.

\subsection{Data Description}

The study selected Android system mobile phones with rich device models and performance that can produce significant differentiation during operation as the original dataset. The data collection process was distributed in four time periods: October and December 2023, April and July 2024, to ensure the impact of seasonal factors on the data. At the same time, objective environmental factors were controlled, and the collection was expanded nationwide. Using the video playback software Douyin as a medium, the performance characteristics of more than 3000 mobile phone models were collected without infringing on user privacy data.

The collection of performance characteristics is implemented internally in the framework as different collectors, each responsible for collecting a second-level category of features, including startup collectors, CPU collectors, memory collectors, power consumption collectors, and profile collectors. Each collector can be configured to collect multiple primary feature items. In the experimental process of the algorithm model, effective device performance profile features were fitted through descriptive time series analysis and principal component analysis methods. At the same time, primary feature items were continuously added or reduced, and the final collected data items were determined to be Table~\ref{tab:1}.

\begin{table}
\centering
\renewcommand{\tablename}{Table}
\caption{Device Performance Feature Data Collection Items}\label{tab:1}
\fontsize{8}{10}\selectfont 
\begin{tabular}{@{}ccc@{}}
\toprule[1.5pt]
	    \textbf{Primary Category.}  & \textbf{Secondary Feature Item}   & \textbf{Data Units} \\ \midrule[1pt]
	 \multirow{2}*{Start}  &cold\_start\_duration               & ms        \\
	                                 &first\_swipe\_duration              &ms    \\
	    		 Play          &first\_frame\_duration               & ms        \\
	 \multirow{4}*{Memory}   & java\_memory\_usage\_ratio             & \%        \\
	                                 & block\_gc\_count             & $-$       \\
	                                 & block\_gc\_time              & ms        \\
	                                 & block\_gc\_per\_time              & ms       \\
          \multirow{2}*{CPU}  & cpu\_usage\_ratio               & \%       \\
                                          & cpu\_speed                  & $GHz$        \\
	\multirow{5}*{Energy}    & battery               & $-$        \\ 
	                                 & battery\_temprature              & $-$        \\
	                                 & power\_saving\_mode               & $-$        \\
	                                 & temprature\_level              & $-$       \\
	                                 & is\_charge               & $-$        \\                           
	\multirow{2}*{Net}   & WiFi             & $-$        \\
	                                & net\_level            & $-$        \\
	\multirow{5}*{UI}     &UI\_frame\_drop\_count            & $-$        \\
	                               &UI\_frame\_drop\_max            & $-$        \\
	                               &UI\_block\_count           & $-$        \\
	                               &UI\_block\_time           & ms        \\
	                               &frame\_rate                               & fps        \\
			Others      & bytebench            & $-$        \\   \\ \bottomrule[1.5pt]
\end{tabular}
\end{table}

The network quality data and device profile data such as heating level in the table are descriptive time series analyzed and aggregated based on device ID and device model. The network quality value ranges from 0 to 8, with higher values indicating better network quality. The heating level value ranges from 0 to 8, with higher values indicating higher real-time heating temperature for the device as a whole. In addition, principal component analysis was performed on numerous secondary feature items within each primary category, and an appropriate number of feature items were selected for collection and experimentation. After preprocessing methods such as missing value and outlier handling and feature availability evaluation, a total of 12 million data points were used for experimental validation.

\subsection{Feature Construction Validation}

Fitting objective and effective performance features or device-level portrait features from collected raw data is an important part of this study. The hierarchical relationship and correlation between features largely affect the evaluation results of performance status. In the process of principal component analysis, determining correlation coefficients, and constructing time series features, all feature items should be input into the secondary evaluation model in modules and levels. The key operations include the following parts.

\subsubsection{Feature Preprocessing}

During the feature value preprocessing, the value range of each feature was determined based on its actual meaning, and data outside the value range was considered as outliers and discarded. Linear interpolation was performed on null and missing values in the time series sequence of the feature at the device level.

\subsubsection{Portrait Feature Construction}

Analyzing and experimentally verifying the construction of network quality portraits as an example.

\begin{figure}
  \centering
  \begin{minipage}[t]{0.5\linewidth}
  \includegraphics[width=1\textwidth]{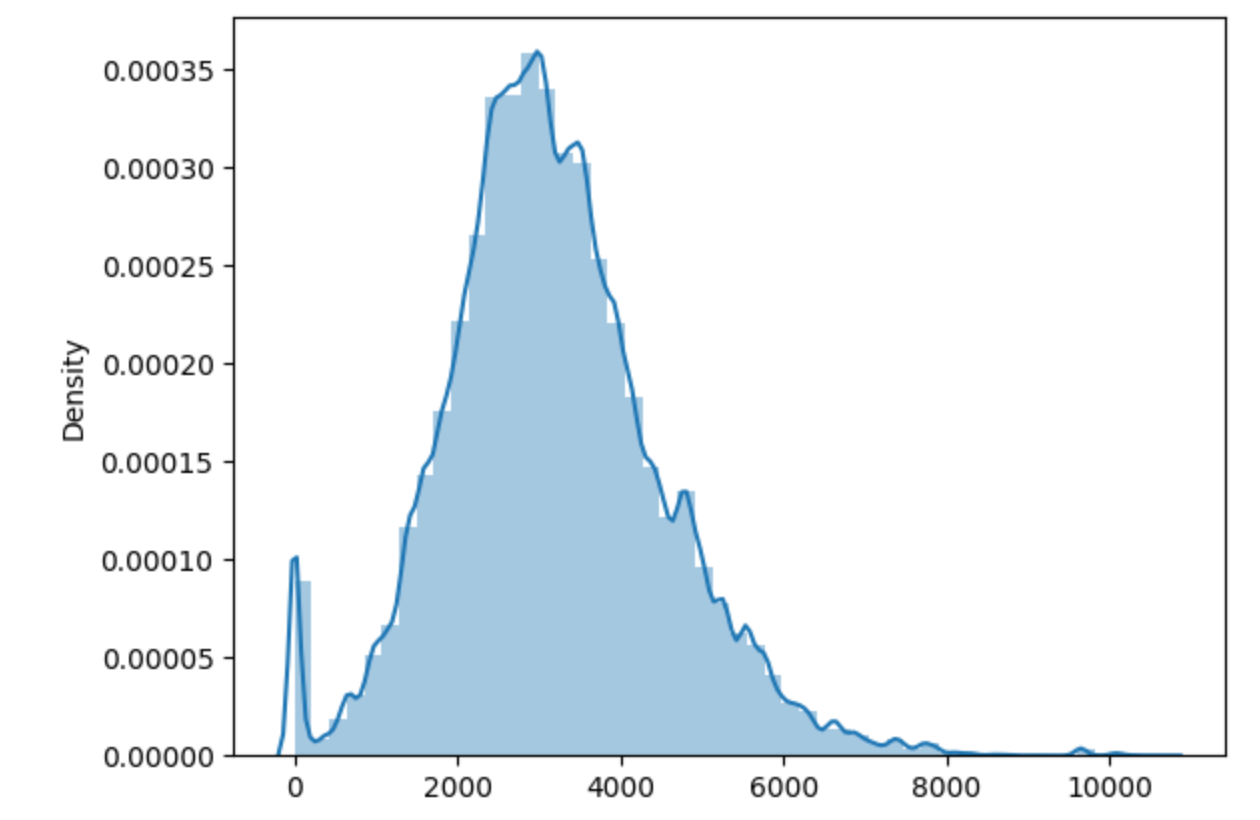}
  \caption{Distribution of 4G network speed values.}
  \label{fig:4}
\end{minipage}%
\begin{minipage}[t]{0.5\linewidth}
  \centering
  \includegraphics[width=1\textwidth]{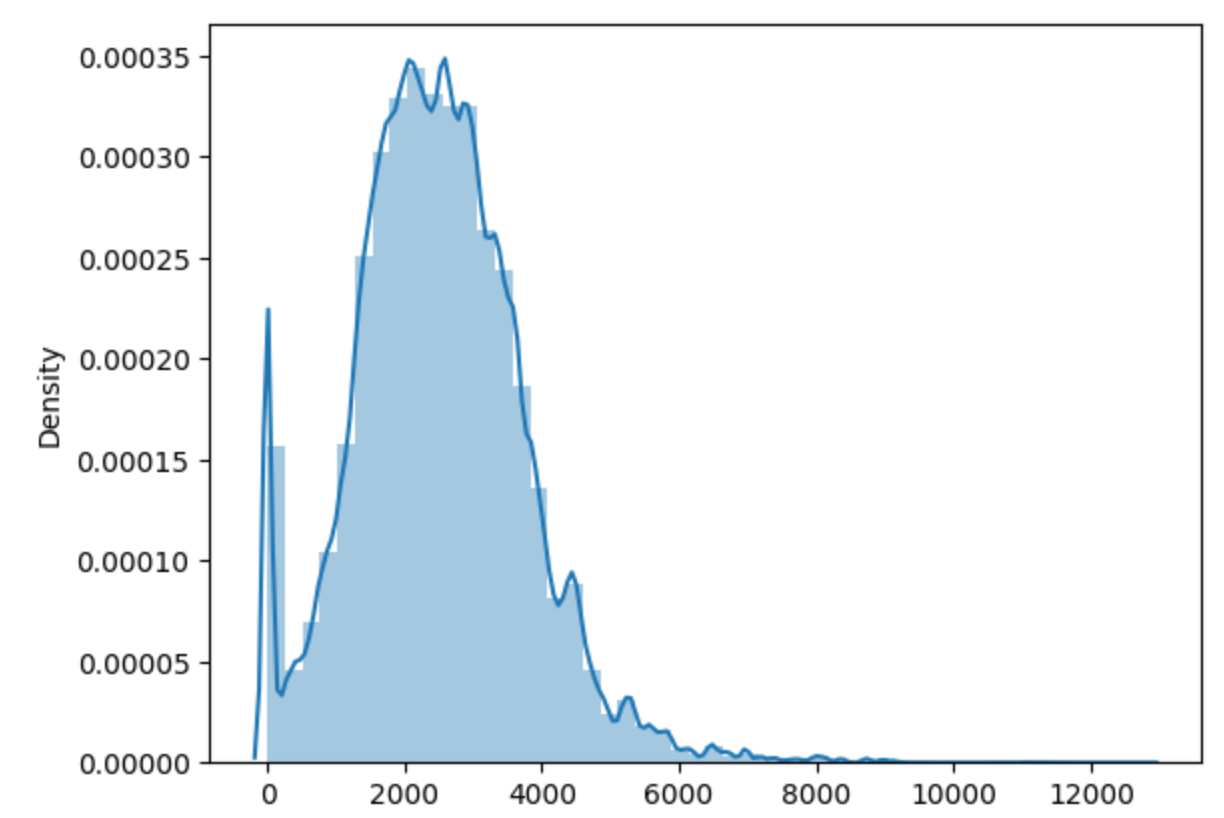}
  \caption{Distribution of Wi-Fi network speed values.}
  \label{fig:5}
  \end{minipage}%
\end{figure}

In the original dataset of 12 million, network-related feature items were collected based on device granularity, and selected based on the correlation coefficient between each feature and the overall network quality: the network speed value of the device in 4G/5G networks, the network speed value of the device in Wi-Fi networks, the duration of the first frame rendering of video playback on the device, and the stuttering rate of the device during use (i.e. video playback behavior). The collected feature item data was analyzed for distribution as shown in Figure~\ref{fig:4} and Figure~\ref{fig:5}, while long and short-term time-domain sequences were constructed to observe the trend of network speed values over time.

The network quality label is defined in the Table~\ref{tab:2} based on the changes in the numerical values of the distributed combined time series.  The logic of portrait fitting tags is shown in the Table~\ref{tab:3} and Table~\ref{tab:4}. 

\begin{table}
\centering
\renewcommand{\tablename}{Table}
\caption{Definition Logic of Network Quality Labels}\label{tab:2}
\fontsize{8}{10}\selectfont 
\begin{tabular}{@{}ccc@{}}
\toprule[1.5pt]
	    \textbf{Tag Category}  & \textbf{Features}   & \textbf{Value Range} \\ \midrule[1pt]
	\multirow{4}*{0}   &4g\_speed  & unpredictable        \\
	                           &wifi\_speed  & unpredictable        \\
	                           &ffd\_avg  & unpredictable        \\
	                           &block\_pct  & unpredictable        \\
	 \multirow{4}*{1}      &4g\_speed  & $\>3000KB/s$         \\
	                               &wifi\_speed  & $\>3500KB/s$        \\
	                           &ffd\_avg  &$<=140ms$        \\
	                           &block\_pct  &  $<=0$       \\
	 \multirow{4}*{2}     &4g\_speed  & $>1600KB/s$ $\&\& <=3000KB/s$        \\
	                               &wifi\_speed  &  $>1600KB/s$ $\&\& <=3500KB/s$        \\
	                           &ffd\_avg  & $>140ms$ $\&\& <=300ms$      \\
	                           &block\_pct   &$>0$ $\&\& <=0.01$        \\
	 \multirow{4}*{3}     &4g\_speed  &$<=1600KB/s$       \\
	                               &wifi\_speed  & $<=1600KB/s$        \\
	                           &4ffd\_avg  & $\>300ms$        \\
	                           &block\_pct  &  $\>0.01$        \\         \\ \bottomrule[1.5pt]
\end{tabular}
\end{table}

\begin{table}
\centering
\renewcommand{\tablename}{Table}
\caption{ Network Quality Portrait Fitting Logic}\label{tab:3}
\fontsize{8}{10}\selectfont 
\begin{tabular}{@{}cc@{}}
\toprule[1.5pt]
	    \textbf{Tag Category}  & \textbf{Independent Variable}    \\ \midrule[1pt]
			0   & \multirow{4}*{Historical distribution of tags in the last 15 days}       \\
	                 1           &                   \\
	                 2            &               \\
	                 3             &             \\
         \\ \bottomrule[1.5pt]
\end{tabular}
\end{table}

\begin{table}
\centering
\renewcommand{\tablename}{Table}
\caption{ Network Quality Portrait Fitting Logic}\label{tab:4}
\fontsize{8}{10}\selectfont 
\begin{tabular}{@{}cc@{}}
\toprule[1.5pt]
	    \textbf{Tag Category}  & \textbf{Establishment Conditions} \\ \midrule[1pt]
			0    &  No available historical tags        \\
	                 1            & The proportion of tags with a label of 1 in the past day and a history of 1 is at least 70\%        \\
	                 2            &  Cases other than 0, 1, and 3        \\
	                 3            & The proportion of tags with a label of 3 in the past day and a history of 3 is at least 70\%       \\
         \\ \bottomrule[1.5pt]
\end{tabular}
\end{table}

The original dataset of 12 million network quality feature data was divided into experimental and test datasets using a 4:1 time domain interval segmentation for fitting. After constructing the time series of the experimental dataset and fitting the labels according to the defined rules, the precision, recall, and other results of the predicted segmentation of the test dataset were verified. The results are shown in Table~\ref{tab:5} and Table~\ref{tab:6}.

\begin{table}
\centering
\renewcommand{\tablename}{Table}
\caption{ Various feature label fitting test accuracy}\label{tab:5}
\fontsize{8}{10}\selectfont 
\begin{tabular}{@{}cccc@{}}
\toprule[1.5pt]
	    \textbf{Predicted Target}  & \textbf{Category}   & \textbf{Prediction Proportion}  & \textbf{Accuracy} \\ \midrule[1pt]
\multirow{2}*{Wi-Fi}   &1  & 13.70\%  & 79.80\%       \\
                             &3  & 6.70\%  & 72.90\%       \\
\multirow{2}*{4G}   &1  & 4.10\%  & 61.60\%        \\
                             &3  & 7.90\%  & 69.50\%        \\
\multirow{2}*{First Frame}   &1  & 5.60\%  & 81.80\%       \\
                             &3  & 9.10\%  & 78.50\%       \\
\multirow{2}*{Block}   &1  & 29.60\%  & 80.50\%       \\
                             &3  & 7.50\%  & 65.60\%       \\                              
         \\ \bottomrule[1.5pt]
\end{tabular}
\end{table}

\begin{table}
\centering
\renewcommand{\tablename}{Table}
\caption{Various feature label fitting test recall}\label{tab:6}
\fontsize{8}{10}\selectfont 
\begin{tabular}{@{}ccccc@{}}
\toprule[1.5pt]
	    \textbf{Predicted Target}  & \textbf{Category}   & \textbf{Recall Proportion} & \textbf{Recall Rate} & \textbf{Stability}\\ \midrule[1pt]
\multirow{2}*{Wi-Fi}   &1     & 10.90\%  & 51.24\%   & 73.47\%    \\
                             &3     & 4.90\%  & 42.20\%   & 71.75\%    \\
\multirow{2}*{4G}   &1     & 2.50\%  & $-$   & $-$     \\
                             &3     & 5.50\%  & $-$    & $-$     \\
\multirow{2}*{First Frame}   &1     & 4.60\%  & 27.20\%   & 72.66\%    \\
                             &3     & 7.10\%  & 45.40\%   & 84.76\%    \\
\multirow{2}*{Block}   &1     & 23.80\%  & 50.00\%   & 78.25\%    \\
                             &3     & 5.00\%  & 30.10\%   & 81.91\%    \\                              
         \\ \bottomrule[1.5pt]
\end{tabular}
\end{table}

By fitting the characteristics of the label and combining them with the above correlation coefficients, the network quality portrait label is obtained as one of the features input into the downstream performance state real-time evaluation module. Similarly, other descriptive time series device portraits such as battery heating level and device model score can be obtained. This objectively describes the performance state of the device and makes up for the shortcomings of similar features that are difficult to obtain, requiring subjective judgment and complex processing.

\subsection{Real-time Performance Evaluation}

\begin{figure}
  \centering
  \begin{minipage}[t]{0.5\linewidth}
  \includegraphics[width=1\textwidth]{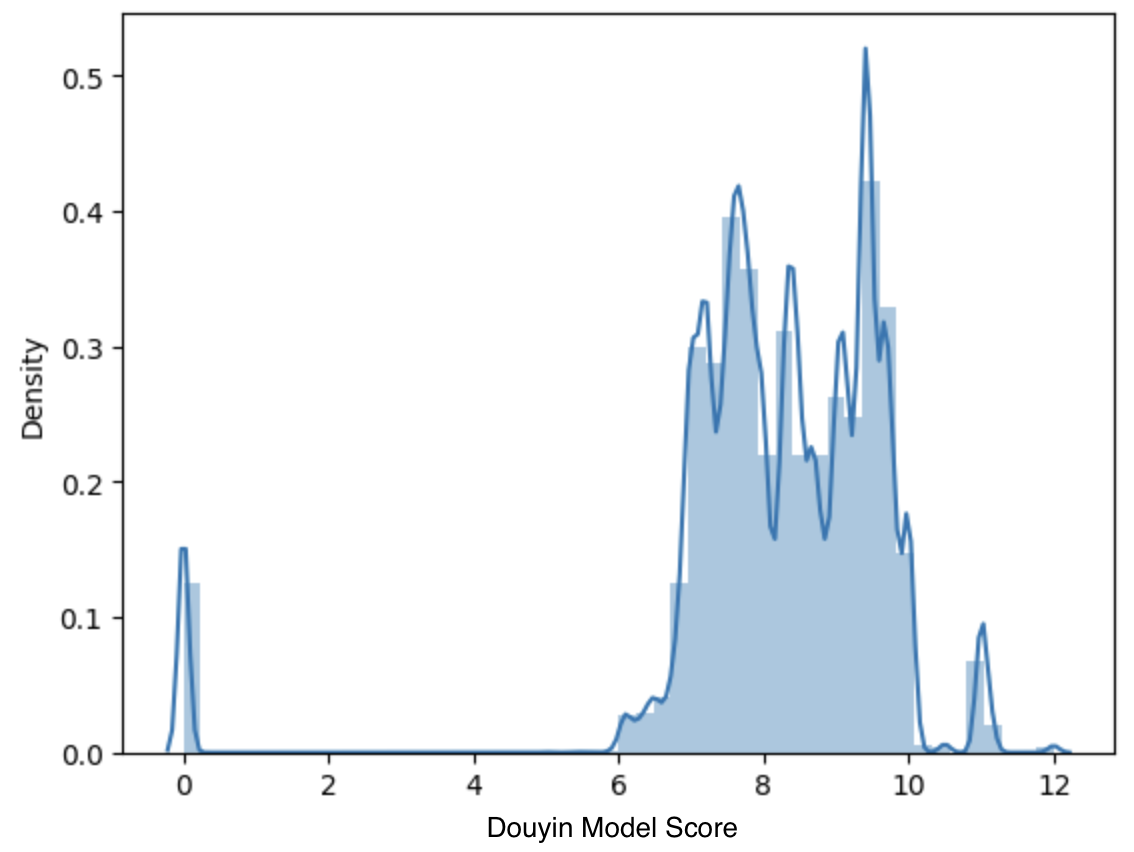}
  \caption{Douyin model performance standardization score distribution density.}
  \label{fig:6}
\end{minipage}%
\begin{minipage}[t]{0.5\linewidth}
  \centering
  \includegraphics[width=1\textwidth]{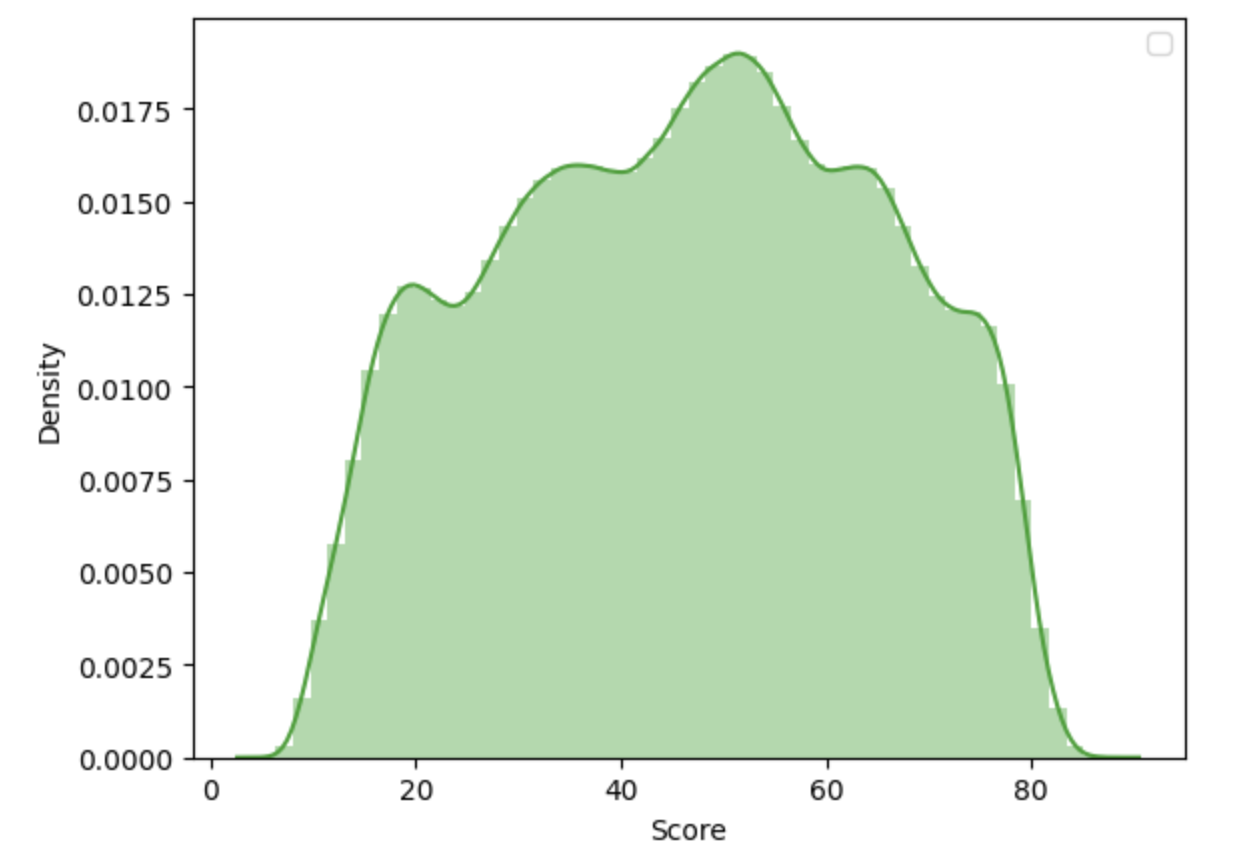}
  \caption{Device real-time performance status score distribution density.}
  \label{fig:7}
  \end{minipage}%
\end{figure}

The algorithm process combining the TOPSIS method with entropy weight method in Part 2 was used to evaluate the performance status of the devices in the original dataset. In order to distinguish the performance differences between different models and devices within the same model, the scoring range for overall performance was set to $[0,100]$ in this experiment. According to the performance rating standards provided by Douyin for different models, which were normalized to $(0,12]$, Figure~\ref{fig:6} shows the distribution density of model performance scores, and Figure~\ref{fig:7} shows the real-time distribution density of performance status ratings.

\begin{figure}
  \centering
  \includegraphics[width=0.9\textwidth]{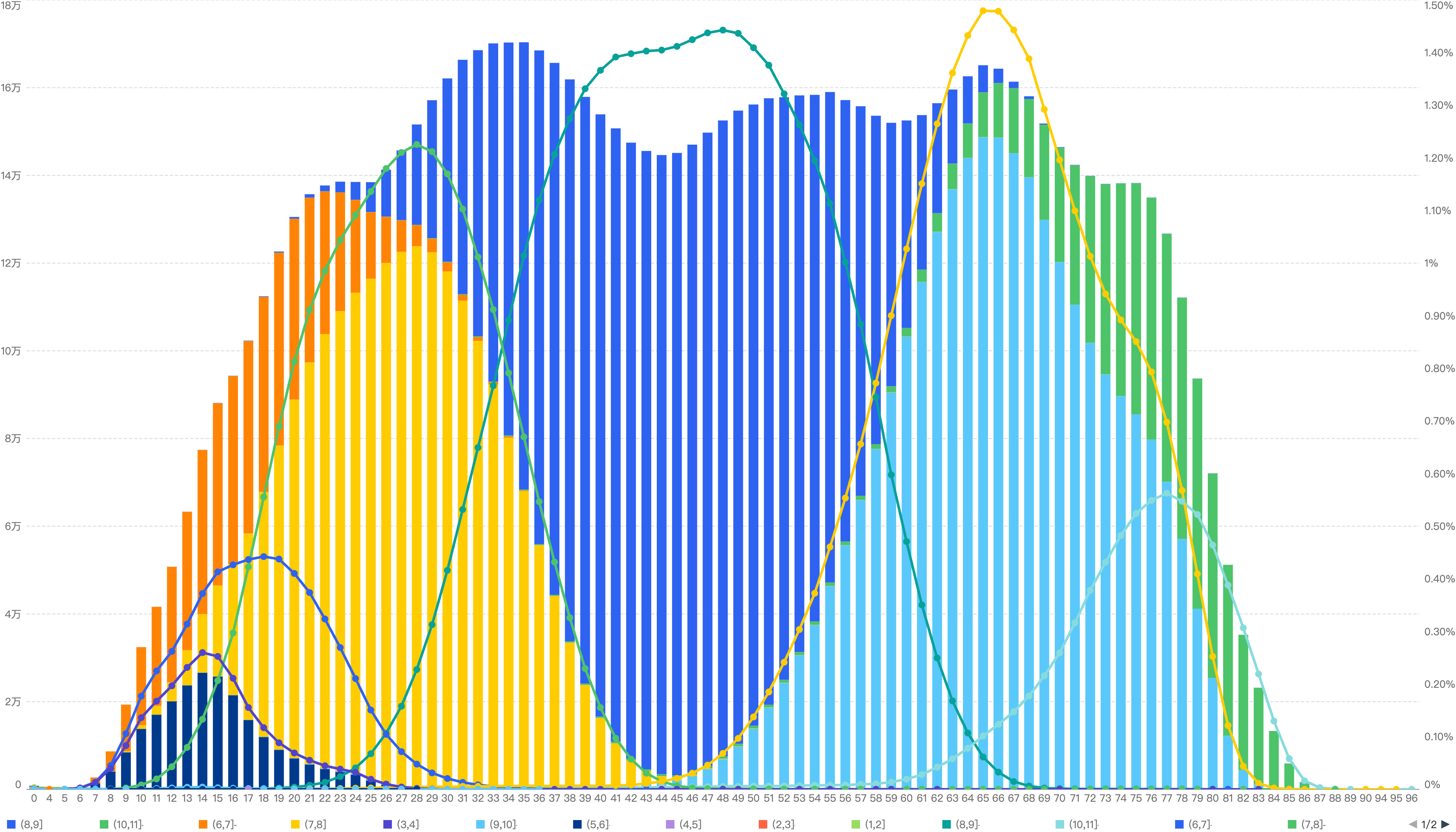}
  \caption{Distribution of dynamic performance of equipment is classified into intervals under standardized device-models.}
  \label{fig:8}
\end{figure}

The two distribution charts show that the performance score distribution of model granularity and the dynamic performance score distribution of device granularity are roughly similar, presenting an overall normal distribution. There are fewer models and devices with extremely poor or excellent performance scores, while there are more models and devices in the middle performance state. As for the difference between model granularity and device dynamic score, it can be visually observed through Figure~\ref{fig:8}. The rectangular bars with the same color in the figure represent devices in the same model score interval, and the horizontal axis score is the dynamic performance score of device granularity.

\begin{figure}
  \centering
  \includegraphics[width=1\textwidth]{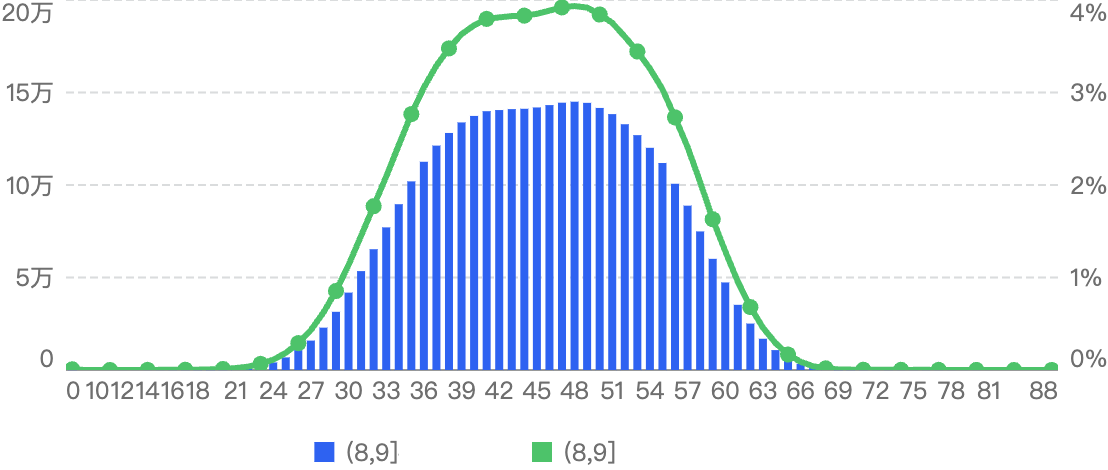}
  \caption{Real-time performance evaluation score distribution for device-models in the $(8,9]$ range.}
  \label{fig:9}
\end{figure}

For example, the distribution chart in Figure~\ref{fig:9} shows that devices with original model ratings in the $(8,9]$ nterval have significant personalized differences in dynamic performance evaluation, with dynamic score thresholds ranging from a minimum of 23 points to a maximum of 76 points. Normalizing the distribution of model ratings in the entire dataset to the $[0,100]$ interval range, the lowest score that can be mapped to the $(8,9]$ model rating is 54.17 points, and the highest score is 76.05 points. This indicates that a more personalized and refined evaluation can be achieved in the evaluation process of dynamic performance. At the same time, the fact that the highest dynamic score is lower than the model rating also confirms the objective fact that the dynamic performance of mobile devices during use is lower than the static performance based on the original model configuration.

Validating the usability and effectiveness of dynamic performance scoring through practical application is a major challenge in this study. By using the numerical threshold values for high, medium, and low-end Android devices as defined by the official video playback software, i.e. low-end threshold $(0,7.21]$, medium-end threshold $(7.21,8.65]$, and high-end threshold $(8.65,12.0)$, the distribution proportions of the three types of devices in the entire raw dataset were normalized to the range of $[0,100]$. The proportions of low-end, medium-end, and high-end devices were $13.45\%$, $39.66\%$, and $46.89\%$, respectively. This resulted in the low-end threshold being $(0,28.67]$, the medium-end threshold being $(28.67,56.82]$, and the high-end threshold being $(56.82,100)$.

\begin{figure}
  \centering
  \includegraphics[width=1\textwidth]{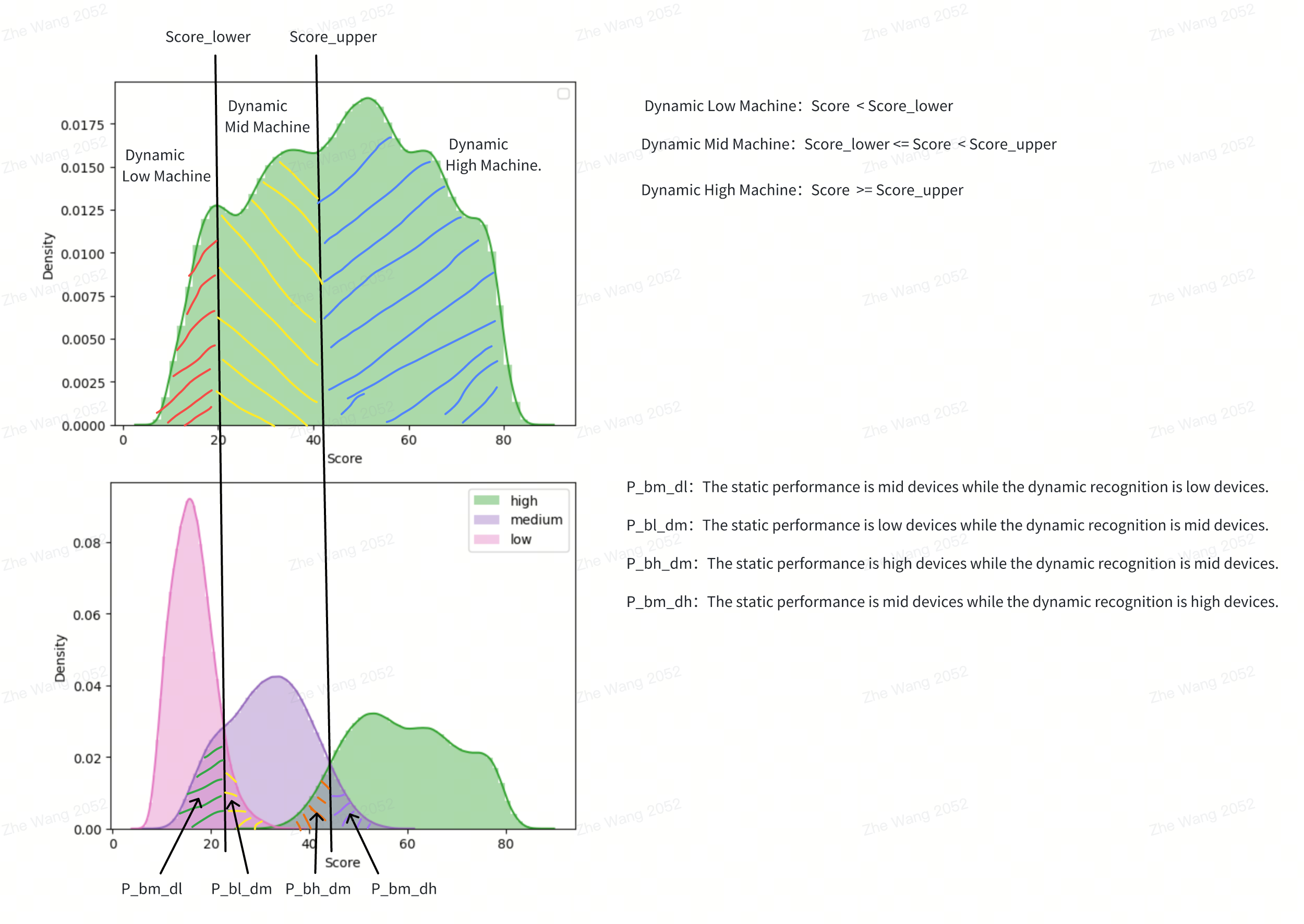}
  \caption{Based on static machine classification distribution, dynamic performance is divided by device.}
  \label{fig:10}
\end{figure}

Firstly, understand how to verify the availability and effectiveness of performance dynamic segmentation through Figure~\ref{fig:10}. In this study, we designed a fine-tuned playback strategy for devices with differences between static machine type segmentation and dynamic performance segmentation to prove that the performance of devices has differentiated decline during video playback. The devices with differences between static machine type segmentation and dynamic performance segmentation are normalized and mapped to the score interval as shown in Figure~\ref{fig:10}. The devices with differences are the intersection of the area chart.

Design an AB experiment based on real-time data of video playback software Douyin. The control group A uses only low-end devices partitioned by static machine type (i.e. devices with machine type scores falling in the interval $(0,7]$ with reduced power consumption strategies, while the experimental group B uses only low-end devices partitioned by dynamic performance scores (i.e. devices with dynamic performance scores falling in the threshold interval $(0,28.67]$ with the same strategy. The number of dynamically assigned devices in the control and experimental groups remained the same throughout the experiment. After a one-month verification period, the degree of performance optimization under the same strategy between the two groups of devices can verify the availability and effectiveness of dynamic performance scores, as shown in Table~\ref{tab:7}.

\begin{table}[htbp]
\centering
\renewcommand{\tablename}{Table}
\caption{Experimental Results}\label{tab:7}
\fontsize{8}{10}\selectfont 
\begin{tabular}{@{}cccc@{}}
\toprule[1.5pt]
	    \textbf{Experimental Group}  & \textbf{Performance Metrics}   & \textbf{Value}    & \textbf{Optimization} \\ \midrule[1pt]
		\multirow{4}*{Control group}   & Playback Stability (ANR average penetration)  & 13.777891     & $-$   \\
		                                    & Playback Smoothness        &17.29ms    & $-$   \\
		                                    & First Brush Time Spent                     &11.654002ms              & $-$    \\
		                                    &  Resource Occupancy              &32\%    & $-$  \\
		\multirow{4}*{Experimental group}   & Playback Stability (ANR average penetration)   & 11.777891     & $-0.443\%$   \\
		                                    & Playback Smoothness         &5.29ms    & $-0.628\%$   \\
		                                    & First Brush Time Spent                      &8.2115409ms              & $-0.108\%$    \\
		                                    & Resource Occupancy               &26\%    & $-6\%$  \\         \\ \bottomrule[1.5pt]
\end{tabular}
\end{table}

\section{Conclusion}
\label{sec:Con}

In the application research of real-time perception of device performance status, we have deeply explored objective evaluation methods for dynamic performance, aiming to solve the key challenges of real-time evaluation of device performance in complex business scenarios. Through comprehensive analysis of relevant algorithms and technologies, we have realized a framework for objectively evaluating the comprehensive performance characteristics of devices that dynamically change during use, and continuously monitoring their performance status. Experimental results show that our method has achieved significant results in both timeliness and accuracy.

Specifically, this article has achieved the following important results in applied research:
\begin{enumerate}
  \item 
By combining descriptive time-series analysis with feature correlation analysis, we fitted performance labels for device profiling. The presentation characteristics of various classification performances of the device over a period of time were objectively described, which compensated for the shortcomings of difficult feature acquisition, complex numerical processing, and subjective description without objective dependence.
  \item 
  A multi-level performance real-time evaluation model based on objective empowerment has been implemented. It can dynamically collect the performance characteristics of devices in real-time and accurately evaluate their performance through TOPSIS under the dimensionality reduction methods of entropy weighting and principal component analysis.
  \item 
  Realized the short-term comprehensive evaluation and prediction of device performance. Based on time series smoothing and prediction model construction, the average performance state of the device within a certain time window is comprehensively evaluated, and the performance state of the next period is predicted, realizing the early perception and warning of device performance and effectively capturing the subtle changes in performance to ensure the stable operation of the device.
\end{enumerate}
In addition, there are still some limitations in this study. In future work, further iterations and improvements can be made. For example, expanding the scope of performance perception to cover more performance aspects and application scenarios to meet more refined business needs; strengthening the integration with deep learning technology to improve the intelligence level of performance perception and prediction; conducting larger-scale practical application verification to better evaluate the effectiveness of this solution and its adaptability in various business scenarios. In summary, this study provides valuable theoretical and practical references for the field of performance dynamic perception, laying a solid foundation for further refinement of device performance and operational stability experience adjustments.

\bibliographystyle{unsrt}  
\bibliography{references}

\begin{thebibliography}{1}

\bibitem{papathanasiou2018topsis}
Jason Papathanasiou, Nikolaos Ploskas, Jason Papathanasiou, and Nikolaos
  Ploskas.
\newblock {\em Topsis}.
\newblock Springer, 2018.

\bibitem{1965Lectures}
Ludwig Boltzmann.
\newblock Lectures on gas theory.
\newblock {\em American Journal of Physics}, 33(11):974--975, 1965.

\bibitem{2018OPTIMISING}
Aistis Raudys and Židrina PABARŠKAITĖ.
\newblock Optimising the smoothness and accuracy of moving average for stock
  price data.
\newblock {\em Technological and Economic Development of Economy},
  24(3):984--1003, 2018.

\bibitem{2014A}
Jonathon Shlens.
\newblock A tutorial on principal component analysis.
\newblock {\em International Journal of Remote Sensing}, 51(2), 2014.

\bibitem{1994A}
Pao~Long Chang and Yaw~Chu Chen.
\newblock A fuzzy multi-criteria decision making method for technology transfer
  strategy selection in biotechnology.
\newblock {\em Fuzzy Sets \& Systems}, 63(2):131--139, 1994.

\bibitem{2000A}
Alberto Montanari, Renzo Rosso, and Murad~S. Taqqu.
\newblock A seasonal fractional arima model applied to the nile river monthly
  flows at aswan.
\newblock {\em Water Resources Research}, 36(5):1249--1259, 2000.

\end{thebibliography}

\end{document}